\documentclass[11pt]{article}

\usepackage{geometry}                		
\usepackage{graphicx}				
							
\usepackage{color}
\usepackage{amsmath,amssymb}
\usepackage[hidelinks]{hyperref}			
\newcommand\blfootnote[1]{
    \begingroup
    \renewcommand\thefootnote{}\footnote{#1}
    \addtocounter{footnote}{-1}
    \endgroup
}
\newcommand{\be}{\begin{equation}}
\newcommand{\ee}{\end{equation}}
\newcommand{\ttb}{{T\bar{T}}}

\newcommand{\CI}{{\mathcal{I}}}

\newcommand{\CO}{{\mathcal{O}}}

\newcommand{\CS}{{\mathcal{S}}}

\newcommand{\p}{\partial}

\newcommand{\half}{\frac{1}{2}}

\newcommand{\Tr}{\mathrm{Tr}\,}
\newcommand{\bk}{\bold{k}}
\begin{document}
\begin{titlepage}
\flushright RCHEP-24-002\\
\hfill \\
\vspace*{15mm}
\begin{center}
{\Large \bf Quantum Complexity of $T\bar{T}$-deformation\\
 and Its Implications}

\vspace*{15mm}

{\large $\text{Amin Faraji Astaneh}^{a,b}\blfootnote{\href{mailto:faraji@sharif.edu}{email: faraji@sharif.ir}}$}
\vspace*{8mm}

\parbox{ \linewidth}{\begin{center}$^a$Department of Physics, Sharif University of Technology,\\
P.O.Box 11155-9161, Tehran, Iran\\ \vspace{0.5cm} 
$^b$Research Center for High Energy Physics\\
Department of Physics, Sharif University of Technology,\\
P.O.Box 11155-9161, Tehran, Iran
\end{center}}\\

\vspace*{0.7cm}

\end{center}
\begin{abstract}
We employ holography to calculate the quantum complexity of $\ttb$-deformation, utilizing the complexity equals volume (CV) and the complexity equals action (CA) proposals within the bulk spacetime with a finite radius cutoff. We find that the complexity of the deformed theory differs from the renormalized complexity of the original theory by a geometric functional: the bending (Willmore) energy of the time-constant slice of the base manifold. We use this result to propose an unambiguous scheme for calculating holographic quantum complexity through the CA proposal.
\end{abstract}

\end{titlepage}

\newpage

\section{Introduction}
The $T\bar{T}$-deformation stands out as a highly significant topic in today's quantum field theory landscape. What makes this deformation particularly intriguing is that it is irrelevant while still being integrable, a unique feature not often seen in conventional field theories \cite{Zamolodchikov:2004ce}-\cite{Bonelli:2018kik}.

The $T\bar{T}$ deformation, introduced in its original form in two dimensions, is constructed by inserting the determinant of the energy-momentum tensor of a given field theory as a composite operator. More precisely, this composite operator is defined as
\be
\CO_\ttb=\half\epsilon^{ij}\epsilon^{k\ell}T^\lambda_{i\ell}T^\lambda_{jk}\, ,
\ee
where $T^\lambda_{ij}$ is the energy momentum tensor of deformed theory, and $\lambda$ denotes the parameter of deformation. The flow equation for the deformed action reads then
\be\label{flow}
\p_\lambda S_\lambda=\int\sqrt{h}\, \CO_\ttb\, .
\ee
One can begin with the action of an original theory and solve the above flow equation to obtain the deformed action in order. 

Interestingly, there are a few useful and simple holographic prescriptions that retain geometric intuition available in correspondence with this deformation. One of them is known as the \emph{holography at finite cut-off} \cite{V}, see also \cite{Guica}-\cite{Taylor}. According to this proposal, the deformation is correspondingly implemented in the bulk by removing the asymptotic region of the bulk spacetime and placing the theory on a finite radius cut-off. The finite radius of the cut-off surface then would be proportionally related to the parameter of deformation. 
By analyzing the energy spectrum of the deformed theory, we can gain insights that guide us toward this holographic proposal. The deformation parameter can indeed take on two opposite signs. We stick to the conventions of \cite{V}, where the positive sign is considered relevant. In this case, the spectrum is bounded from below; however, it becomes imaginary at very high undeformed energies. To avoid this, one must remain sufficiently distant from the UV modes, which are located very close to the boundary of AdS. This leads us to eliminate the asymptotic region of the bulk space by imposing a finite radius cut-off. In parallel, and considering the opposite sign, a single-trace deformation of string theories in AdS$_3$ has been introduced, which exhibits many aspects of the $\ttb$-deformation. For details, one can refer to \cite{Asrat:2017tzd}-\cite{Giveon:2017nie}, and we also suggest \cite{Jiang:2019epa} for a general discussion on the signs of the deformation parameter.

This finite cut-off proposal has been verified in many aspects. One of the primary verifications is the coincidence of the energy spectrum in both field theory and gravity sides. 
See also  \cite{Bhattacharyya:2023gvg}-\cite{He:2024xbi} for other recent developments in this area.

This proposal also has provided the capability to compute the various quantities of deformed theory in the context of quantum information theory. For example, one can follow the holographic proposals for computing the entanglement entropy and replace the asymptotic boundary with a finite radius cut-off. Specifically, one can analytically calculate the entanglement entropy for the deformed theory defined on a sphere, with the entangling region being a hemisphere. The holographic proposal introduced above is capable of replicating this result in the context of AdS/CFT correspondence \cite{Donnelly:2018bef}. For further developments in this area, see \cite{Park:2018snf}-\cite{FarajiAstaneh:2022qck}. A key object in such calculations is an extremal holographic hypersurface.  A subtle point to note is that the original theory lives on the asymptotic boundary and we only renormalize the extremal volumes, while the modified theory lives on a finite boundary from the beginning and the extremal volumes are all automatically renormalized.

In this study, we aim to investigate another informative quantity, the quantum complexity of deformation. This measure will inform us how complex it is to deform a given quantum state, by $\ttb$-insertion. To perform this calculation, we utilize two well-known holographic proposals, complexity equals volume and complexity equals action. One could refer to \cite{Chandra:2021kdv} to \cite{Roy2} for related studies.

In particular, \cite{Chandra:2021kdv} attempts to establish a general holographic framework that relates circuit complexity to gravitational action. As the authors of this paper propose, one can think of the AdS geometry as describing a sequence of states that are interconnected through Euclidean time evolution and changes in the UV cut-off. In this context, there is an evolution between two states at time $t_i$ and radial coordinate $z_i$ to the time $t_f$ and radial coordinate $z_f$. This evolution will define a finite cut-off with a profile $z = \rho(t)$, and the authors suggest that the gravitational action in the region enclosed by this non-constant radius cut-off represents the circuit complexity of the associated field theory.  
Focusing on two dimensions, this analysis finds connections with the path integral optimization proposal, then, finding the profile of the finite cut-off surface that minimizes the complexity would be a relevant endeavor addressed in this paper. Nevertheless, we will continue to use the CA complexity proposal when computing the gravitational action, and we will seek to establish an appropriate scheme of regularization in generic $d$-dimensions, aiming to uncover some geometric notions of complexity related to $\ttb$-deformation.

We will employ these holographic proposals in a bulk space with finite cut-off with the most generality to finally get a geometric expression for the complexity of deformation on the boundary. We ultimately arrive at a geometric interpretation of the quantum complexity associated with $\ttb$-deformation, using which we propose a scheme for computing complexity through the complexity equals action proposal, which has faced ambiguities for a while.

Before going ahead, it is important to emphasize that the holographic setup provides a general framework that is expected to work in $d$-dimensions without any subtleties. We thus will work in general $d$-dimensions while still using the abbreviation $\ttb$, which was originally conceived for a two-dimensional field theory, in a broader context. 

This paper is organized as follows: In Section 2, we calculate the complexity of deformation using the complexity equals volume proposal. In section 3 we express our geometric intuition. Then in section 4, we repeat the calculation following another holographic recipe, the complexity equals action proposal. In the last part, we conclude this study.

\section{Complexity equals volume}
In the first step, we compute the quantum complexity utilizing a well-known holographic proposal the so-called complexity equals volume (CV) proposal \cite{Susskind:2014rva}-\cite{Susskind:2014moa} (see also \cite{Alishahiha:2015rta} for the complexity of the sub-regions). According to this proposal, the complexity is determined in terms of the volume of a holographic minimal hypersurface which is governed by extending a time-constant slice of the boundary into the bulk. Denoting this minimal hypersurface by $\Sigma$ the complexity reads then\footnote{It is necessary to include a dimensionful length scale in the denominator, which is commonly taken to be the radius of AdS. However, we have chosen to set this length scale to unity.}
\be
C_V=\mathrm{Min}\frac{V_\Sigma}{G_N}\, .
\ee
We have discussed the setup in appendix \eqref{app1}, in which we have clarified the notation, and provided detailed explanations on how to appropriately write a metric describing the setup. We will adopt the holographic setup in its broadest form as outlined in \cite{Carmi:2016wjl}.

We start with the Fefferman-Graham form of the bulk metric in the following form
\be\label{metric}
\begin{split}
ds^2_N&=G_{\mu\nu}dX^\mu dX^\nu\\
&=\frac{d\rho^2}{4\rho^2}+\frac{1}{\rho}\left\{-(1+\rho S_{ nn})dt^2+\left[h_{ab}+2 t  K_{ab}+ t^2(K^2_{ab}-R_{anbn})-\rho S_{ab}\right]dy^ady^b\right\}\, ,
\end{split} 
\ee
where the radius of AdS is set to unity. $K_{ab}$ is the extrinsic curvature of the time-constant slice of the base manifold, $\sigma$ and $S_{ij}$ is $d$-dimensional Schoutten tensor defined as \eqref{Schoutten}.
Tensors with index $n$ denote the contractions with the unit normal on $\sigma$.
In this expansion, we have kept terms up to order $\CO(t^2)$ and $\CO(\rho)$.

For our calculations, we need the expansion of the volume element in the bulk
\be\label{VE}
\begin{split}
&\sqrt{G}=\frac{\sqrt{h}}{2\rho^{-\frac{d+2}{2}}}\sum_{m,n=0} V^{(m,n)}(x)t^m\rho^n\, ,\\
\end{split}
\ee
The first few terms of this expansion read (see appendix \eqref{app1})
\be
\begin{split}
&V^{(1,0)}=\frac{1}{2}\Tr g^{(1,0)}=K\, , \\
&V^{(2,0)}=\frac{1}{8}\left[4\Tr g^{(2,0)}-2\Tr {g^{(1,0)}}^2+(\Tr g^{(1,0)})^2\right]=-\frac{1}{2}(\Tr K^2-K^2+R_{nn})\, ,\\
&V^{(0,1)}=\frac{1}{2}\Tr g^{(0,1)}=-\frac{1}{2}S^i_i=-\frac{1}{4(d-1)}R\, .
\end{split}
\ee
What we aim to calculate is the volume of a holographic minimal hypersurface, $\Sigma$.
This hypersurface is parametrized as
\be
 t = t(\rho,y)=\sum_{k=1} \tau_k(y)(\rho-\rho_c)^k\, ,
\ee
and the coefficients $\tau_k(y)$ will be fixed with the demand of extrimization of the hypersurface $\Sigma$, i.e.
\be
K_{\Sigma}=\frac{1}{\sqrt{G}}\p_\mu (\sqrt{G}G^{\mu\nu}N_\nu)=0\ ,
\ee
where $N_{\mu}$ denotes the normal vector on $\Sigma$. Using \eqref{VE} and
solving the above equation order by order one finds
\be
\tau_1=\frac{1}{2(d-1)}V^{(1,0)}\, ,
\ee
which is enough for our purposes. Using this expansion of the embedding function one would be able to determine the induced metric on minimal hypersurface $\Sigma$. The details of the calculation are presented in appendix \eqref{IM}.
Doing so, the volume element on extremal $\Sigma$ takes the following expansion
\be
\sqrt{H}=\frac{1}{2\rho^{\frac{d+1}{2}}}[1+\half\rho S^a_a+(\rho-\rho_c)K\tau_1+\cdots]\, .
\ee
where
\be\label{SctoEi}
S^a_a=\frac{1}{d-2}\left(R^a_a-\half R\right)=\frac{1}{d-2}\left(R_{nn}+\half R\right)\, .
\ee
and we have used $R^a_a=R+R_{nn}$. The last term gives the normal-normal component of the Einstein tensor $G_{nn}=R_{nn}+\half R$.
One can now evaluate the volume of $\Sigma$ and then the CV complexity as follows
\be\label{CVd}
\begin{split}
C_V^{(\ttb)}&=\frac{1}{G_N}\int d^dY\sqrt{H}=\frac{1}{G_N}\int d^{d-1}y\int_{\rho_c}d\rho\sqrt{H}\\
&=\frac{\rho_c^{\frac{1-d}{2}}}{(d-1) G_N}\left[V_\sigma-\frac{(d-1)\rho_c}{2(d-3)(d-2)}\int_\sigma d^{d-1}y\sqrt{h}\left(G_{nn} -\frac{d-2}{(d-1)^2}K^2\right)\right]\, .
\end{split}
\ee
It should be emphasized that the finite cut-off radius is related to the parameter of deformation, specifically as $\rho_c\propto 1/\lambda$, as previously mentioned \cite{V} and \cite{FarajiAstaneh:2024fig}.

Interestingly enough, the Gauss-Codazzi identity gives
\be\label{GC}
G_{nn}=\half R_\sigma-\half(\Tr K^2-K^2)\, ,
\ee
and thus $C_V^{(\ttb)}$ can be entirely written in terms of the Euler number of $\sigma$ and a functional of its extrinsic curvature.

It is important to note that the complexity for the original theory reads \cite{Carmi:2016wjl}, 
\be\label{CVo}
\begin{split}
C_V^{(0)}&=\frac{\rho_c^{\frac{1-d}{2}}}{(d-1)G_N}\left[V_\sigma-\frac{(d-1)\rho_c}{2(d-3)}\int_\sigma d^{d-1}y\sqrt{h}\left(S^a_a -\frac{(d-2)}{(d-1)^2}K^2\right)\right]\\
&=\frac{\rho_c^{\frac{1-d}{2}}}{(d-1)G_N}\left[V_\sigma-\frac{(d-1)\rho_c}{2(d-3)(d-2)}\int_\sigma d^{d-1}y\sqrt{h}\left(G_{nn} -\frac{(d-2)^2}{(d-1)^2}K^2\right)\right]\, .
\end{split}
\ee
Therefore
\be\label{dCV}
\delta C_V\equiv C_V^{(0)}-C_V^{(\ttb)}=\frac{\rho_c^{\frac{3-d}{2}}}{G_N}W_\sigma\, ,
\ee
where
\be
W_\sigma=\frac{1}{2(d-1)^2}\int_\sigma d^{d-1}y\sqrt{h}\, K^2\, ,
\ee
is the bending (Willmore) energy of hypersurface $\sigma$. In the following section, we will provide a more precise introduction to this quantity and discuss the physical intuition that supports this result.
Before that, it is worth commenting on the special case of $d=2$ dimensions. In two dimensions, the Einstein tensor vanishes, and the integral of $K^2$ drops out of the complexity calculated for the undeformed theory, as shown in \eqref{CVo}. Consequently, the subleading contribution to the complexity of the undeformed theory appears at the next order. In contrast, the Willmore energy of the time-constant slice of the two-dimensional base manifold does contribute to the complexity, as indicated in \eqref{CVd}. This causes the complexity of deformation to be present in any dimension as indicated above, since nontrivial cancellations of factors in the denominator of the subleading term in \eqref{CVo} occur due to the presence of $K^2$ term in the complexity of the two-dimensional deformed theory\footnote{We would like to thank the referee for pointing out this issue.}. 
\section{Willmore energy and the physical intuition about the complexity of $\ttb$-deformation}
The key idea of the holographic prescription for $\ttb$-deformation, namely holography at a finite cut-off, is the holographic realization of the deformed theory within a finite radius cut-off of AdS spacetime. From a geometric perspective, it is crucial to account for the bending energy associated with this finite radius cut-off. This energy is quantified by the Willmore functional of the hypersurface, which is defined as the square of the trace of the extrinsic curvature of the hypersurface, integrated along the finite cut-off boundary \cite{W}, \cite{FarajiAstaneh:2024fig}.

The fact that by introducing this bending term, one can holographically generate the action of the deformed theory is the subject of \cite{FarajiAstaneh:2024fig}. And now, this geometric term has reemerged in the calculation of the quantum complexity of deformation.

To explain our result, let us consider a quantum computer as a collection of correlated quantum gates, for which we define two tasks. The first task is to generate the original theory and then to renormalize it. The second task is to generate the deformed theory on the finite cut-off from the beginning. Our findings indicate that the second task is simpler, and the difference between the two complexities of the tasks is quantifiable as the bending energy of the finite cut-off hypersurface. We will verify this finding through the calculation of the complexity using another well-known holographic prescription, the so-called complexity equals action, in the following section.

\section{Complexity equals action}
In this proposal, the hypersurface $\sigma$ serves as the boundary of a specific region within spacetime, known as the Wheeler-DeWitt (WDW) patch. The WDW patch represents the domain of dependence for Cauchy surfaces within the bulk, aligning asymptotically with the time-constant slice of the asymptotic boundary. According to the CA proposal, the complexity of the state is expressed as \cite{Brown:2015bva}, \cite{Brown:2015lvg}, and \cite{Lehner:2016vdi}
\be
C_A = \frac{S_{WDW}}{\pi \hbar}\, ,
\ee
where, $S_{WDW}$ denotes the gravitational on-shell action computed within the WDW patch, including all boundary terms. In the following subsections, the evaluation of each part of the action will be presented individually.
\subsection{Bulk term}
The WDW patch is identified as a section of AdS enclosed by two future/past null boundaries, symbolized as, $\CI_+$ and $\CI_-$, respectively. We collectively denote the null boundaries of the WDW patch as $\Sigma$. By definition, we have (\cite{Carmi:2016wjl}; see also \cite{Solodukhin:2008qx}).
\begin{equation}
\CI_\pm\, :\, t=t_\pm(\rho,y)\, , \, \frac{\p t_\pm}{\p\rho}=\pm\frac{\sqrt{g_{\rho\rho}}}{\sqrt{-g_{tt}}}=\frac{1}{2}(\rho^{-\frac{1}{2}}-\frac{1}{2}S_{nn}\rho^{\frac{1}{2}})\, .
\end{equation}
Therefore
\be
t_\pm(\rho,y)=\pm(\rho^{\frac{1}{2}}-\rho_c^{\frac{1}{2}})\mp\frac{1}{6}S_{nn}(\rho^{\frac{3}{2}}-\rho_c^{\frac{3}{2}})\, .
\ee
The Einstein-Hilbert action simply measures the volume of the WDW patch
\be
S_{EH}=\frac{1}{16\pi G_N}\int_{WDW}d^{d+1}X\sqrt{G}\, (R_G-2\Lambda)=-\frac{d}{8\pi G_N}V_{WDW}\, ,
\ee
which up to the few leading terms reads
\be\label{EH2}
\begin{split}
S_{EH}&=-\frac{d}{8\pi G_N}\int d^{d-1}y\sqrt{h}\int_{\rho_c} d\rho\int_{t_-}^{t_+}dt\left(1+\half \rho S_{nn}\right)\\
&\times\frac{1}{2\rho^{\frac{d+2}{2}}}\left(1+V^{(1,0)}t+V^{(2,0)}t^2+\hat{V}^{(0,1)}\rho\right)\, .
\end{split}
\ee
Here in $\hat{V}^{(0,1)}$ the trace is constructed using the induced metric on $\sigma$. More accurately
\be
\hat{V}^{(0,1)}=\half h^{ab}g^{(0,1)}_{ab}=-\half S^a_a=-\frac{1}{2(d-2)}G_{nn}\, ,
\ee
where we have used \eqref{SctoEi}. 

Performing the integrals one finally finds\footnote{The authors of \cite{Carmi:2016wjl} use a similar renormalization setup, but they obtain a different result for the subleading term, as indicated in equation (3.11) of their paper. In investigating the source of the mismatch, we noticed that the overall sign of equation (D.12) might be negative; with this adjustment, one arrives at $G_{nn}$ in the integrand of the bulk action. This may help explain why (3.11) behaves differently in dimensions $d=3$ and $d=7$ which is unexpected. Of course, the authors have examined their formula in Appendix C of their paper for a particular geometry; however, according to (C.17) in that paper, the right-hand side of (D.12) vanishes and therefore the potentially problematic term does not contribute to the result of the particular example.}
 \be\label{EH}
\begin{split}
S_{EH}=-\frac{\rho_c^{\frac{1-d}{2}}}{4(d-1)\pi G_N}\left[V_\sigma-\frac{\rho_c}{(d-3)(d-2)}\int_\sigma d^{d-1}y\sqrt{h}\left(\Tr K^2-K^2+\frac{d+1}{2}G_{nn}\right)\right]\, ,
\end{split}
\ee
Thanks to the Gauss-Codazzi identity \eqref{GC}, $S_{EH}$ can be entirely written in terms of the Euler number of $\sigma$ and a functional of its extrinsic curvature.
\subsection{Joint term}
This action should be accompanied by some boundary terms. 
A very important point that we will take advantage of, is that in the holographic picture of deformed theory, the only asymptotic boundary is located where future/past null boundaries meet. Therefore a joint null-null boundary term is all that is required to ensure the well-posedness of the variational principle. Such a boundary term has been thoroughly explored in \cite{Lehner:2016vdi}. This joint term is constructed from the null normals at the location of $\sigma$, as follows
\be
S_j=-\frac{1}{8\pi G_N}\int d^{d-1}y\sqrt{H}\,\log\big\vert \frac{\bk_+\cdot \bk_-}{2}\big\vert_{\rho=\rho_c}\, ,
\ee
where
\be
\bk_\pm=\alpha_\pm (k_\pm^\rho,k_\pm^t,k_\pm^a)\, ,
\ee
represent the future/past null normals, respectively. Here, $\alpha_\pm$ is an undetermined normalization constant. Up to the leading orders in $\rho$, one finds
\be
\begin{cases}
k_\pm^\rho=-2\rho^{3/2}(1-\frac{1}{2}S_{nn}\rho)\, ,\\
k_\pm^t=\mp\rho(1-S_{nn}\rho)\, ,\\
k_\pm^a=\mp S_n^a\rho^2\, ,
\end{cases}
\ee
where upon using the Gauss-Codazzi identity, $S_n^a=n^tS_t^a$ includes the intrinsic derivatives of the extrinsic curvature on $\sigma$. Therefore, one ultimately obtains, up to leading orders
\be\label{Ij}
\bk_+\cdot\bk_-=2\alpha_+\alpha_-\rho(1-S_{nn}\rho)\, ,
\ee
and thus
\be
\log\big\vert \frac{\bk_+\cdot \bk_-}{2}\big\vert_{\rho=\rho_c}\simeq \log(\alpha_+\alpha_-\rho_c)-S_{nn}\rho_c\, .
\ee
The appearance of unwanted undetermined normalization constants may seem pathological; however, there is no need for a particular concern. As we will see shortly, these terms can be covariantly eliminated by choosing appropriate counterterms.
\subsection{Counterterms}
We have encountered two unpleasant features thus far. Firstly, the leading term of the bulk action makes a negative leading contribution to the complexity, and secondly, there is an ambiguity due to the dependence on the undetermined normalization constants of the null vectors in the joint term. Our aim in introducing the counterterms is to address these two problems with minimal intervention. Therefore, we add the counterterms with the following two aims:
\begin{itemize}
\item
To resolve the ambiguity through a minimal subtraction.
 \item
To recover the volume law while ensuring the positivity of complexity.
\end{itemize}
We assert that these minimal demands define an appropriate scheme for the computation of the complexity. As we will see, the outcome is consistent with our intuitive expectations and reassures us that this is a suitable scheme for the holographic calculation of the quantum complexity through the CA proposal. We deal with these counterterms in two following parts, individually.

Let us begin with our primary aim. As previously mentioned, there is no need to be concerned about the unwanted dependence on undetermined normalization constants since it would be removed by introducing the following counterterm \cite{Reynolds:2016rvl}\footnote{This counterterm was originally introduced in \cite{Lehner:2016vdi} but as indicated in \cite{Reynolds:2016rvl} a more minimal counterterm would fulfill our objective of eliminating ambiguities arising from the undetermined normalization constants in null vectors. In fact, the counterterm introduced above results from an integration by parts of the more general counterterm presented in \cite{Lehner:2016vdi}. We have found it crucial to adopt such a minimal subtraction. See also \cite{Akhavan:2019zax} for a related investigation of the counterterms.}
\be
S_{c.t.}^{(1)}=\frac{1}{8\pi G_N}\int d^{d-1}y\sqrt{H}\,\log\left(\frac{\Theta}{d-1}\right)\, ,
\ee
one both future and past null boundaries, where $\Theta$ denotes the expansion of null generators with the following definition
\be
\Theta=\frac{1}{\sqrt{H}}\frac{\p}{\p\lambda}\sqrt{H}\, ,
\ee
with $\lambda$ serving as the affine parameter on the null boundaries. Henceforth, we will express everything individually on the future and past null boundaries, avoiding the encapsulated formulas. The affine parameter would be identified through the definition $k^\mu_\pm=\frac{\p X^\mu}{\p\lambda_\pm}$. This up to the leading orders yields
\be
\lambda_\pm=\mp\frac{1}{\alpha_\pm}\rho^{-1/2}\left(1-\frac{1}{2}S_{nn}\rho\right)\, .
\ee
Returning to the definition of $\Theta$, one explicitly finds
\be\label{ITheta}
\begin{split}
\Theta_\pm&=\alpha_\pm(d-1)\rho^{1/2}\Big[1\mp\frac{1}{2(d-1)}\Tr g^{(1,0)}\rho^{1/2}\\
&+\frac{1}{2(d-1)}\rho^{1/2}\left(\rho^{1/2}-\rho_c^{1/2}\right)({\Tr g^{(1,0)}}^2-\Tr g^{(2,0)})\\
&-\frac{1}{d-1}\rho\left(\widehat{\mathrm{Tr}}\, g^{(0,1)}+\frac{d-1}{2}S_{nn}\right)\Big]\, ,
\end{split}
\ee
where $\widehat{\mathrm{Tr}}$ denotes the trace calculated using the intrinsic metric $h^{ab}$, as given in Equation \eqref{EH2}.
Therefore,
\be
\begin{split}
&\log\left(\frac{\Theta_+}{d-1}\right)+\log\left(\frac{\Theta_-}{d-1}\right)-\log\left(\frac{\bk_+\cdot\bk_-}{2}\right)\\
&\simeq-\frac{\rho_c}{d-1}\left[\frac{1}{d-1}K^2-2S^a_a\right]\simeq\frac{2\rho_c}{(d-2)(d-1)}G_{nn}-2\rho_c W_\sigma\, .
\end{split}
\ee
and thus
\be\label{jct}
S_j+S_{c.t.}^{(1)}=\frac{\rho_c^{\frac{3-d}{2}}}{4\pi G_N}\frac{1}{(d-2)(d-1)}\int d^{d-1}y\, \sqrt{h}\, G_{nn}-\frac{\rho_c^{\frac{3-d}{2}}}{4\pi G_N}W_\sigma\, .
\ee
We would like to highlight the elimination of $S_{nn}$ between the integrands in \eqref{Ij} and \eqref{ITheta}. This will clean up the scene, leading to a more elegant final result, which ultimately supports our assertions regarding quantum complexity in the context of the CA proposal.

As evidenced by \eqref{jct}, the resulting boundary term is still subleading in the finite radius cut-off parameter. Therefore, we are still confronted with a negative leading volume term in the Einstein-Hilbert bulk action, \eqref{EH}. On the other hand, we intuitively expect that the leading term of the complexity exhibits a volume law dependence, much like in CV complexity, \eqref{CVd}. The reasoning is simply that at leading order a larger construction is inherently expected to accommodate greater complexity. Much more information will be captured in the subleading terms. Therefore, it is reasonable to expect that the complexity scales with the volume in leading order. Interestingly enough, we can recover such a volume dependence at the leading order by introducing the standard counterterm commonly used in the holographic reconstruction of spacetime \cite{SSD}. By incorporating this counterterm, we ultimately arrive at a positive complexity that features a leading volume term. We will explore it in detail through the following steps.

Parallel to what is commonly done in holographic renormalization, we add to the whole integrand a suitable coefficient of the volume element on the null boundaries and at the asymptotic limit, acting as a covariant counterterm. For us, this is $\frac{1}{4\pi(d-1)G_N}\sqrt{H}$ along the null boundaries and when the radial coordinate in the bulk tends to $\rho_c$. This gives us the following contribution to the boundary terms
\be
S_{c.t.}^{(2)}=\frac{\rho_c^{\frac{1-d}{2}}}{2\pi(d-1)G_N}-\frac{\rho_c^{\frac{3-d}{2}}}{4\pi G_N}\frac{1}{(d-2)(d-1)}\int d^{d-1}y\, \sqrt{h}\, G_{nn}\, .
\ee
Notably, the second term cancels the first term in \eqref{jct} which aligns with our expectation to minimally touch the action while adding the required boundary terms. On the other hand, it is worth mentioning that the first term when combined with the leading term of the Einstein-Hilbert action, provides a positive volume-like leading contribution to the complexity, as desired.

Therefore putting things together, one finally gets\footnote{Henceforth, we set $4\pi^2\hbar=1$ to align the prefactors with CV complexity..}
\be
\begin{split}
&C_A^{(\ttb)}=4\pi(S_{EH}+S_j+S_{c.t.}^{(1)}+S_{c.t.}^{(2)})\\
&=\frac{\rho_c^{\frac{1-d}{2}}}{(d-1) G_N}\left[V_\sigma-\frac{\rho_c}{2(d-3)(d-2)}\int_\sigma d^{d-1}y\sqrt{h}\left(2\Tr K^2-2K^2+(d+1)G_{nn}\right)\right]\\
&-\frac{\rho_c^{\frac{3-d}{2}}}{G_N}W_\sigma\, .
\end{split}
\ee
This is the final result for the CA complexity of deformation and here we would like to make the following comments in order:
\begin{itemize}
\item The complexity is positively defined as it should be and the leading term demonstrates a volume-law as desired. We have achieved these two advantages, by adding some minimal counterterms covariantely. To the best of our knowledge, such a volume-dependent leading term in CA complexity has not been addressed in prior works.
\item The second term in the parenthesis above is also of interest. It has been written in terms of the projected Einstein tensor as well as scalars constructed from the extrinsic curvature tensor of $\sigma$. This combination resembles a similar term in CV complexity, \eqref{CVd}. Utilizing the Gauss-Codazzi identity, \eqref{GC} it can be recast in terms of the Euler term and the projected Einstein tensor, or alternatively, the Euler term and the scalars made of the extrinsic curvature tensors of $\sigma$.
\item Just similar to CV complexity, the bending (Willmore) energy of $\sigma$ is singled out in CA complexity. It is worth mentioning that this singling out arises naturally from the minimal required boundary terms we added by inquiring about some basic demands rather than being deliberately extracted from the action. This observation strengthens our ideas about the complexity of deformation, suggesting that the complexity of the deformed theory differs from the normalized complexity of the original undeformed theory by the amount of the Willmore energy evaluated on $\sigma$.
\item The unique form of the CV and CA complexities for deformed theory encourages us to take our proposal regarding the complexity of deformation and its relation to the complexity of renormalized original theory more seriously. According to this idea which has been verified with our findings, the difference between the complexities of a deformed theory living on the finite cut-off boundary and the renormalized original theory amounts to the Willmore energy of the hypersurface on which we evaluate the complexity. It is worth highlighting that the holographic calculation of CA complexity for deformed theory is much clearer than for the original theory. In the latter case, the WDW patch includes a segment of the asymptotic boundary, where various, possibly different, boundary terms can be imposed. This scheme ambiguity is not concerned with standard holographic renormalization of theories, as it leads to finite terms that are usually disregarded. Nevertheless, in our case, these finite terms are meaningful, representing perturbations in the parameter of deformation. Therefore, a clear scheme of calculation is essentially needed.

We propose that the correct scheme is obtained by evaluating the action in the WDW patch that intersects the finite cut-off boundary. It naturally needs to impose the usual minimal boundary terms. Then one needs just to subtract the bending energy of the hypersurface of constant time to obtain the CA complexity of undeformed theory. In this way, we define a unique scheme for calculating CA complexity. To the best of our knowledge, this is a novel proposal in this regard.
\end{itemize}
Based on the last comment, we propose the following expression for the CA complexity of the original undeformed theory
\be
\begin{split}
&C_A^{(0)}=\frac{\rho_c^{\frac{1-d}{2}}}{(d-1) G_N}\left[V_\sigma-\frac{\rho_c}{2(d-3)(d-2)}\int_\sigma d^{d-1}y\sqrt{h}\left(2\Tr K^2-2K^2+(d+1)G_{nn}\right)\right]\, ,\\
\end{split}
\ee
and thus
\be
\delta C_A\equiv C_A^{(0)}-C_A^{(\ttb)}=\frac{\rho_c^{\frac{3-d}{2}}}{G_N}W_\sigma\, ,
\ee
which exactly matches what we already get for CV complexity, \eqref{dCV}.
Now we are ready to conclude our study.
\section{Conclusion}
In this note, we have investigated the quantum complexity of $\ttb$-like deformations. We have employed the established holographic proposals, specifically the complexity equals volume and the complexity equals action, in our computations. Our bulk setup is designed to reveal the $\ttb$-deformed theory at the boundary. As is well known, this is accomplished by removing the asymptotic region in the bulk and introducing a finite cutoff boundary. Our investigation follows a general survey in $d$-dimensions which has been explored in some aspects of undeformed theories, previously.

The critical step in this computation is to employ a holographic setup with a finite radius cutoff, as mentioned above. Accordingly one may expect a renormalized functional of intrinsic and extrinsic covariant scalars on a time-constant slice of the base manifold as the quantum complexity of deformed theory. This expectation is indeed fulfilled, but getting a renormalized expression is not the entire story. We have explicitly demonstrated that the $\ttb$-deformation yields a structure analogous to the renormalized complexity of the undeformed theory. However, there is a notable difference due to the bending (Willmore) energy of the time-constant slice of the base manifold located at the finite cutoff boundary. In this sense, one could conclude that deforming a theory with 
$\ttb$-insertion is simpler than renormalizing a theory, originally defined on the asymptotic boundary (at infinity or zero radius, depending on the choice of the coordinate system).

Interestingly, this observation may be explained by the superluminal propagation speed of the degrees of freedom (relative to the fixed background metric) inherent in a theory deformed by $\ttb$-insertion, \cite{V} and \cite{Kraus:2018xrn}. These superluminal degrees of freedom could facilitate a faster circuit, which might account for the difference in complexity observed between the deformed and undeformed theories. We leave a more detailed investigation to uncover a deeper connection in this regard for future studies\footnote{We would like to thank the referee for suggesting to think in this direction.}.

Remarkably, establishing the setup for calculating the complexity of the deformed theory through the CA proposal is simpler and clearer in the case of deformed theory. 
This is because the WDW patch corresponding to the deformed theory intersects the finite cutoff boundary at a joining hypersurface of co-dimension one, where we finally calculate the complexity. In this case, the only term needed to ensure that the variational principle is well-posed is a joint term, which is entirely known.
In contrast, renormalizing the original theory, which is holographically governed by cutting a segment of the WDW patch, introduces a time-like boundary in addition to the joint terms, on which one may impose various covariant boundary terms. In this sense, the calculation of the CA complexity for the deformed theory is less ambiguous and simpler.

Therefore, when the original theory is concerned, we propose calculating the CA complexity using the first setup and simply adding the bending energy of the time-constant slice of the base manifold. This approach provides an unambiguous and intuitively relevant scheme for calculating the CA complexity.

This study can be generalized to include fields of various spins in the bulk, as well as to gravitational theories with higher derivative contributions from the curvature tensor. Furthermore, one can examine theories that enjoy different space-time or internal symmetries, such as non-relativistic field theories. Investigating subregion complexities would also be an interesting problem, which we leave for future studies.
\newpage
\appendix
\section{Setup}\label{app1}
Our conventions for setting up the problem are as follows. 
The undeformed CFT lives on a $d$-dimensional manifold $\mathcal{M}_d$, which is characterized by the coordinates $x^i = (t, y^a)$, where $t$ represents time. It is assumed that $\mathcal{M}_d$ serves as the asymptotic boundary of the bulk space $\mathcal{N}_{d+1}$, with $\rho$ serving as the additional extra coordinate. The coordinates of the bulk space are collectively denoted by $X^\mu$. 
We focus on a time-constant slice on $\mathcal{M}_d$, denoted as $\sigma_{d-1}$, covered by the coordinates $y^a$. This surface is distinguished by its unit normal, $n^i = \delta^i_t$, which represents a future-directed spacelike vector. The extension of $\sigma_{d-1}$ into the bulk is denoted as $\Sigma_d$, with coordinates $Y^A = (\rho, y^a)$. The metrics on these manifolds are denoted by $g_{ij}$, $G_{\mu\nu}$, $h_{ab}$, and $H_{AB}$, respectively. We chose the Fefferman-Graham metric for the bulk spacetime
\begin{equation}
ds_N^2 = G_{\mu\nu}dX^\mu dX^\nu = \frac{d\rho^2}{4\rho^2} + \frac{1}{\rho}g_{ij}(x,\rho)dx^idx^j,
\end{equation}

where $g_{ij}(x,\rho)$ admits the following expansion

\begin{equation}
g_{ij} = \sum_{m,n=0} g_{ij}^{(m,n)}\, t^m\rho^n.
\end{equation}

The first terms simply read

\begin{equation}
g_{tt}^{(0,0)} = -1, \quad g_{ab}^{(0,0)} = h_{ab}, \quad g_{ta}^{(0,0)} = 0.
\end{equation}

The next term is determined in terms of the intrinsic curvature of the base manifold as well as the extrinsic curvature of the co-dimension one time constant slice of it. Specifically, since the extrinsic curvature reads

\begin{equation}
K_{ij} = \frac{1}{2}\left.\frac{\partial g_{ij}}{\partial t}\right|_{t=0,\rho=0},
\end{equation}

we conclude that

\begin{equation}
g_{ab}^{(1,0)} = 2K_{ab}.
\end{equation}

On the other hand, the calculation of the Riemann tensor on the base manifold yields

\begin{equation}
R_{atbt} = -\frac{\partial K_{ab}}{\partial t} + K_{ac}K^c_b\, ,
\end{equation}

using that one arrives at

\begin{equation}
g_{ab}^{(2,0)} = -R_{atbt} + K^2_{ab}\, ,
\end{equation}

where $K^2_{ab}$ stands for $K_{ac}K^c_b$.

The next term, which will be found through the holographic reconstruction of spacetime, is the Schouten tensor on the asymptotic boundary of AdS$_{d+1}$

\begin{equation}\label{Schoutten}
g_{ij}^{(0,1)} = -S_{ij} = -\frac{1}{d-2}\left(R^{(0)}_{ij} - \frac{1}{2(d-1)}R^{(0)}g^{(0)}_{ij}\right)\, .
\end{equation}

Putting all these together leads to the metric as mentioned in \eqref{metric}.

\section{Induced metric on the holographic hypersurface of co-dimension one}\label{IM}
The induced metric on hypersurface $\Sigma$ of co-dimension one takes the following form
\be
\begin{split}
ds^2_\Sigma&=H_{AB}dY^AdY^B\\
&=\left[\frac{1}{4\rho^2}+G_{tt}(\p_\rho  t)^2\right]d\rho^2+\left(G_{ab}+G_{tt}\p_a t \p_b t\right)dy^a dy^b+\left(2G_{tt}\p_\rho t \p_a  t\right)\,d\rho dy^a\, .
\end{split}
\ee
In order to determine the determinant of $H_{AB}$ and thus the volume element on $\Sigma$
we firstly define the Schur complement of the $\rho\rho$ component as follows
\be
\CS_{ab}=\left(G_{ab}+G_{tt}\p_a t  \p_b  t +2G_{ab}\p_b  t -\frac{1}{H_{\rho\rho}}G_{tt}^2(\p_\rho  t)^2\p_a  t \p_b  t \right)\Big\vert_{ t = t (\rho,y)}\, ,
\ee
then
\be
\sqrt{H}=\sqrt{H_{\rho\rho}}\sqrt{\CS}\, .
\ee
where
\be\label{rhorho}
\sqrt{H_{\rho\rho}}=\frac{1}{2\rho}\sqrt{1+4\rho^2G_{tt}(\p_\rho t )^2}\, ,
\ee

\end{document}